\documentclass[aps,prd,twocolumn,nofootinbib]{revtex4-1}

\usepackage{graphicx}
\usepackage{hyperref}
\usepackage{epsfig}
\usepackage{slashed}

\begin{document}

\title{Production of the $P$-Wave Excited $B_c$-States through the $Z^0$ Boson Decays}

\author{Zhi Yang}
\author{Xing-Gang Wu}
\email{wuxg@cqu.edu.cn}
\author{Li-Cheng Deng}
\author{Jia-Wei Zhang}
\author{Gu Chen}
\affiliation{Department of Physics, Chongqing University, Chongqing 400044, People's Republic of China}

\date{\today}

\begin{abstract}
In Ref.\cite{z0s}, we have dealt with the production of the two color-singlet $S$-wave $(c\bar{b})$-quarkonium states $B_c(|(c\bar{b})_{\bf 1}[^1S_0]\rangle)$ and $B^*_c(|(c\bar{b})_{\bf 1}[^3S_1]\rangle)$ through the $Z^0$ boson decays. As an important sequential work, we make a further discussion on the production of the more complicated $P$-wave excited $(c\bar{b})$-quarkonium states, i.e. $|(c\bar{b})_{\bf 1}[^1P_1]\rangle$ and $|(c\bar{b})_{\bf 1}[^3P_J]\rangle$ (with $J=(1,2,3)$). More over, we also calculate the channel with the two color-octet quarkonium states $|(c\bar{b})_{\bf 8}[^1S_0]g\rangle$ and $|(c\bar{b})_{\bf 8}[^3S_1]g\rangle$, whose contributions to the decay width maybe at the same order of magnitude as that of the color-singlet $P$-wave states according to the naive nonrelativistic quantum chromodynamics scaling rules. The $P$-wave states shall provide sizable contributions to the $B_c$ production, whose decay width is about $20\%$ of the total decay width $\Gamma_{Z^0\to B_c}$. After summing up all the mentioned $(c\bar{b})$-quarkonium states' contributions, we obtain $\Gamma_{Z^0\to B_c} =235.9^{+352.8}_{-122.0}$ KeV, where the errors are caused by the main uncertainty sources. \\


\end{abstract}

\maketitle

With the luminosity raises up to ${\cal L}\propto 10^{34}cm^{-2}s^{-1}$ or higher as programmed by the Internal Linear Collider \cite{ilc}, i.e. the so called Gigaz \cite{gigaz}, and by the newly purposed $Z$-factory \cite{wjw}, it will open new opportunities not only for high precision physics in the electro-weak sector, but also for hadron physics. The discovery of $B_c$ meson by the Collider Detector at Fermilab \cite{cdf} is one of the important discoveries in heavy quark physics. Due to its particular nature, the $B_c$ meson has attracted wide attention. Its hadronic production and decay properties have been throughly studied by the literature, a minireview of its recent improvements can be found in Ref.\cite{yellow}. For example, its semi-leptonic decays can provide a platform to check the color-octet mechanism of Non-Relativistic Quantum Chromodynamics (NRQCD) theory \cite{bcdecay}.

Recently, we have made a detailed discussion on the production of the spin-singlet $B_c$ ($|(c\bar{b})_{\bf 1}[^1S_0]\rangle$) and the spin-triplet $B^*_c$ ($|(c\bar{b})_{\bf 1}[^3S_1]\rangle$) mesons in Ref.\cite{z0s}, which are produced through the $Z^0$ boson decays and are dealt with under the `New Trace Amplitude Approach'. The higher excited states, such as the $P$-wave states or more strictly the higher excited $(c\bar{b})$-quarkonium states, may directly or indirectly (in a cascade way) decay to the ground state with almost 100\% possibility via electromagnetic or hadronic interactions. So the production of the higher excited $(c\bar{b})$-quarkonium states can be regarded as additional sources of $B_c$ production. The hadronic production of the $P$-wave $(c\bar{b})$-quarkonium states through the dominant gluon-gluon fusion mechanism have been studied in Refs.\cite{p1,p2,p3,p4,p5,p6}. Especially, one can conveniently generate the hadronic $S$-wave and $P$-wave $(c\bar{b})$-quarkonium events by using the generator BCVEGPY \cite{p6}. More over, the indirect production of the $P$-wave $(c\bar{b})$-quarkonium states through the top quark decays have been discussed in Refs.\cite{tbc2,tbc3}. It has been found that the higher excited states as $P$-wave states can provide sizable contribution in both the hadronic production and its indirect production through top quark decays at LHC. So as a compensation, it would be interesting to study the production of the $P$-wave $(c\bar{b})$-quarkonium events through the $Z^0$ decays.

In the NRQCD framework \cite{nrqcd,hqg}, a heavy quarkonium is considered as an expansion of various Fock states. The relative importance among those infinite ingredients is evaluated by the velocity scaling rule. Namely the physical state of $h_{B_c}$ and $\chi^J_{B_c}$ can be decomposed into a series of Fock states as follows,
\begin{equation}
|h_{B_c}\rangle = {\cal O}(v^{0})|(c\bar{b})_{\bf 1}[^{1}P_{1}]\rangle +{\cal O}(v^1)|(c\bar{b})_{\bf 8}[^{1}S_{0}]g\rangle  + \cdots  \label{eq:1}
\end{equation}
and
\begin{equation}
|\chi^J_{B_c}\rangle = {\cal O}(v^{0})|(c\bar{b})_{\bf 1}[^{3}P_{J}]\rangle +{\cal O}(v^1)|(c\bar{b})_{\bf 8}[^{3}S_{1}]g\rangle + \cdots, \label{eq:2}
\end{equation}
where $v$ is the relative velocity, and the symbol $\cdots$ means even higher Fock states. We use the symbols $h_{B_c}$ and $\chi_{B_c}^J$ to denote the four physical $P$-wave states, i.e. $h_{B_c}$ denotes the $P$-wave state with the dominant color-singlet state $(c\bar{b})_{\bf 1}[^1P_1]$ and $\chi_{B_c}^J$ denotes the $P$-wave states with the dominant color-singlet states $(c\bar{b})_{\bf 1}[^3P_J]$ with $J=(1,2,3)$ respectively. Here the thickened subscript of $(c\bar{b})$-quarkonium denotes the color index, ${\bf 1}$ for color-singlet and ${\bf 8}$ for color-octet. As a full estimation of the $P$-wave production, we shall discuss the production of the following Fock states simultaneously, $|(c\bar b)_{\bf 1}[^{1}P_{1}]\rangle$, $|(c\bar b)_{\bf 1}[^{3}P_{J}]\rangle$, $|(c\bar b)_{\bf 8}[^{1}S_{0}] g\rangle$ and $|(c\bar b)_{\bf 8}[^{3}S_{1}] g\rangle$.

\begin{figure}
\includegraphics[width=0.40\textwidth]{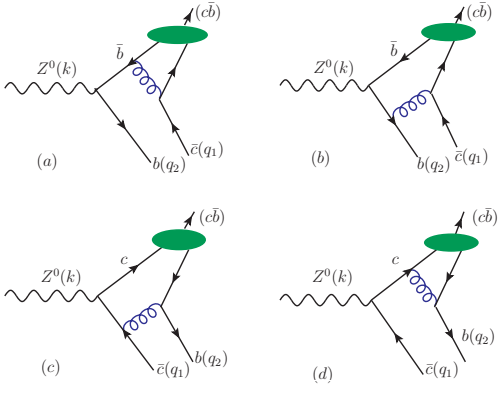}
\caption{Feynman diagrams for the process $Z^0(k)\rightarrow (c\bar{b})(q_3) + b(q_2) +\bar c(q_1)$, where $(c\bar{b})$-quarkonium stands for the Fock states $|(c\bar b)_{\bf 1}[^{1}S_{0}]\rangle$, $|(c\bar b)_{\bf 1}[^{3}S_{1}] \rangle$, $|(c\bar b)_{\bf 1}[^{1}P_{1}]\rangle$, $|(c\bar b)_{\bf 8}[^{1}S_{0}] g\rangle$, $|(c\bar b)_{\bf 1}[^{3}P_{J}]\rangle$ and $|(c\bar b)_{\bf 8}[^{3}S_{1}] g\rangle$ respectively. } \label{feyn}
\end{figure}

The Feynman diagrams for the process $Z^0 \to (c\bar{b}) +b +\bar{c}$ are presented in Fig.(\ref{feyn}), where the $(c\bar{b})$-quarkonium stands for the Fock states $|(c\bar b)_{\bf 1}[^{1}S_{0}]\rangle$, $|(c\bar b)_{\bf 1}[^{3}S_{1}] \rangle$, $|(c\bar b)_{\bf 1}[^{1}P_{1}]\rangle$, $|(c\bar b)_{\bf 8}[^{1}S_{0}] g\rangle$, $|(c\bar b)_{\bf 1}[^{3}P_{J}]\rangle$ and $|(c\bar b)_{\bf 8}[^{3}S_{1}] g\rangle$ respectively. According to the NRQCD factorization formula, the decay width of the process can be written in the following form \cite{nrqcd}
\begin{equation}
d\Gamma=\sum_{n} d\hat\Gamma(Z^0 \to (c\bar{b})[n]+b+\bar{c})\langle{\cal O}^H(n) \rangle,
\end{equation}
where the non-perturbative matrix element $\langle{\cal O}^{H}(n)\rangle$ is proportional to the inclusive transition probability of the perturbative state $(c\bar{b})[n]$ into bound state. The color-octet matrix elements are smaller than the color-singlet matrix elements by certain $v^2$ order. More specifically, based on the velocity scaling rule and under the vacuum-saturation approximation, we have \cite{nrqcd,p5,bcdecay}
\begin{eqnarray}
&&\langle (c\bar{b})_{\bf 8}[^1S_0]|{\cal O}_{\bf 8}(^1S_0)| (c\bar{b})_{\bf 8}[^1S_0] \rangle \nonumber\\
&\simeq&\Delta_S(v)^2\cdot \langle (c\bar{b})_{\bf 1}[^1S_0]|{\cal O}_{\bf 1}(^1S_0)| (c\bar{b})_{\bf 1}[^1S_0] \rangle
\end{eqnarray}
and
\begin{eqnarray}
&& \langle (c\bar{b})_{\bf 8}[^3S_1]|{\cal O}_{\bf 8}(^3S_1)| (c\bar{b})_{\bf 8}[^3S_1] \rangle \nonumber\\
&\simeq& \Delta_S(v)^2\cdot \langle (c\bar{b})_{\bf 1}[^3S_1] |{\cal O}_1(^3S_1)| (c\bar{b})_{\bf 1}[^3S_1] \rangle\,,
\end{eqnarray}
where $\Delta_S(v)$ is of order $v^2$, and we take it to be within the region of [0.10, 0.30], which is consistent with the identification: $\Delta_S(v) \sim \alpha_s(m_{B_c} v)$. For the color-singlet components, the matrix elements can be directly related with the wave functions at the origin for the $S$-wave states or with the first derivative of the wave functions at the origin for the $P$-wave states, which can be computed via the potential models \cite{pot1,pot2,pot3,pot4,pot5,pot6} and/or potential NRQCD (pNRQCD) \cite{pnrqcd} and/or lattice QCD \cite{hqg,lat1}, and references therein.

The short-distance decay width $d\hat\Gamma(Z^{0}\to (c\bar{b})[n] +b +\bar{c})$ can be written as
\begin{equation}
d\hat\Gamma(Z^{0}\to (c\bar{b})[n]+b+\bar{c})= \frac{1}{2k^0} \overline{\sum}  |M|^{2} d\Phi_3,
\end{equation}
where $\overline{\sum}$ means that we need to average over the spin states of the initial particles and to sum over the color and spin of all the final particles. In $Z^0$ rest frame, the three-particle phase space can be written as
\begin{displaymath}
d{\Phi_3}=(2\pi)^4 \delta^{4}\left(k_0 - \sum_f^3 q_{f}\right)\prod_{f=1}^3 \frac{d^3{\vec{q}_f}}{(2\pi)^3 2q_f^0}.
\end{displaymath}
The detailed process for dealing with the $1 \to 3$ phase space can be found in the Appendix A of Ref.\cite{z0s}, from which we can conveniently obtain the differential decay widths such as $d\Gamma/ds_1$, $d\Gamma/ds_2$, $d\Gamma/d\cos\theta_{13}$ and $d\Gamma/d\cos\theta_{23}$, where $s_1=(q_1+q_3)^2$, $s_2=(q_1+q_2)^2$, $\theta_{13}$ is the angle between $\vec{q}_1$ and $\vec{q}_3$, and $\theta_{23}$ is the angle between $\vec{q}_2$ and $\vec{q}_3$.

The hard scattering amplitude for the process $Z^0(k)\rightarrow (c\bar{b})[n](q_3) + b(q_2) +\bar c(q_1)$ can be written as:
\begin{equation}
iM = {\cal{C}}{\bar u_{s i}}({q_2})\sum\limits_{n = 1}^4 {{\cal A} _n }{v_{s' j}}({q_1}), \label{MM}
\end{equation}
where ${\cal{C}}=\frac{e g_s^2}{\sin\theta_{w}\cos\theta_w}\times \frac{4}{3\sqrt{3}}\delta_{ij}$ for the color-singlet case and ${\cal{C}}=\frac{e g_s^2}{\sin\theta_{w}\cos\theta_w}\times (\sqrt{2}T^aT^bT^a)_{ij}$ for the color-octet case ($\sqrt{2}T^b$ stands for the color of the color-octet $(c\bar{b})$-quarkonium state) respectively. The gamma structure ${\cal A}_n$ ($n=1$, $\cdots$, $4$) corresponds to the four Feynman diagrams in Fig.(\ref{feyn}), respectively. ${\cal A}_n$ for the color-singlet $S$-wave states can be found in Ref.\cite{z0s}, and for the color-octet $S$-wave states, one only need to change the color-singlet color factor there to the present color-octet one. While for the $P$-wave states, ${\cal A}_n$ can be written as
\begin{widetext}
\begin{eqnarray}
{\cal A}^{S=0,L=1}_1 &=& \varepsilon_l^{\mu}(q_3) \frac{d}{dq_\mu} \left[ {{\slashed{\epsilon}(k)}{\Gamma_{z\bar b}}\frac{\slashed{q}_2 - \slashed{k} + {m_b}}{(q_2 - k)^2 - m_b^2}{\gamma_\rho} \frac{\Pi^0_{q_3}(q)} {(q_{31} + {q_1})^2}{\gamma_\rho}}\right]_{q=0}, \label{A1}\\
{\cal A}^{S=0,L=1}_2 &=& \varepsilon_l^{\mu}(q_3) \frac{d}{dq_\mu} \left[\gamma_{\rho}\frac{\slashed{k}-\slashed{q}_{32}+{m_b}}{(k-q_{32})^2- m_b^2}{\slashed{\epsilon}(k)}{\Gamma_{z\bar b}} \frac{\Pi^0_{q_3}(q)} {(q_{31} + {q_1})^2}{\gamma_\rho}\right]_{q=0}, \label{A2}\\
{\cal A}^{S=0,L=1}_3 &=& \varepsilon_l^{\mu}(q_3) \frac{d}{dq_\mu} \left[\gamma_{\rho}\frac{\Pi^0_{q_3}(q)}{({q_{32}}+ {q_2})^2} \slashed{\epsilon}(k) \Gamma_{zc} \frac{\slashed{q}_{31} - \slashed{k} + m_c}{(q_{31} - k)^2- m_c^2}{\gamma_\rho}\right]_{q=0}, \label{A3} \\
{\cal A}^{S=0,L=1}_4 &=& \varepsilon_l^{\mu}(q_3) \frac{d}{dq_\mu} \left[\gamma_{\rho}\frac{\Pi^0_{q_3}(q)}{(q_{32} + {q_2})^2} \gamma_{\rho}\frac{\slashed{q}_3 + \slashed{q}_2 + {m_c}}{({q_3} +{q_2})^2 -m_c^2} {\slashed{\epsilon}(k)}{\Gamma_{zc}}\right]_{q=0} \label{A4}
\end{eqnarray}
and
\begin{eqnarray}
{\cal A}^{S=1,L=1}_1 &=& \varepsilon^{J}_{\mu\nu}(q_3) \frac{d}{dq_\mu} \left[ {{\slashed{\epsilon}(k)}{\Gamma_{z\bar b}}\frac{\slashed{q}_2 - \slashed{k} + {m_b}}{(q_2 - k)^2 - m_b^2}{\gamma_\rho} \frac{\Pi^\nu_{q_3}(q)} {(q_{31} + {q_1})^2}{\gamma_\rho}}\right]|_{q=0}, \label{A5}\\
{\cal A}^{S=1,L=1}_2 &=& \varepsilon^{J}_{\mu\nu}(q_3) \frac{d}{dq_\mu} \left[\gamma_{\rho}\frac{\slashed{k}-\slashed{q}_{32}+{m_b}}{(k-q_{32})^2- m_b^2}{\slashed{\epsilon}(k)}{\Gamma_{z\bar b}} \frac{\Pi^\nu_{q_3}(q)} {(q_{31} + {q_1})^2}{\gamma_\rho}\right]|_{q=0}, \label{A6}\\
{\cal A}^{S=1,L=1}_3 &=& \varepsilon^{J}_{\mu\nu}(q_3) \frac{d}{dq_\mu} \left[\gamma_{\rho}\frac{\Pi^\nu_{q_3}(q)}{({q_{32}}+ {q_2})^2} \slashed{\epsilon}(k) \Gamma_{zc} \frac{\slashed{q}_{31} - \slashed{k} + m_c}{(q_{31} - k)^2- m_c^2}{\gamma_\rho}\right]|_{q=0}, \label{A7} \\
{\cal A}^{S=1,L=1}_4 &=& \varepsilon^{J}_{\mu\nu}(q_3) \frac{d}{dq_\mu} \left[\gamma_{\rho}\frac{\Pi^\nu_{q_3}(q)}{(q_{32} + {q_2})^2} \gamma_{\rho}\frac{\slashed{q}_3 + \slashed{q}_2 + {m_c}}{({q_3} +{q_2})^2 -m_c^2} {\slashed{\epsilon}(k)}{\Gamma_{zc}} \right]|_{q=0}, \label{A8}
\end{eqnarray}
\end{widetext}
where $\Gamma_{z\bar b} =\frac{1}{4} - \frac{1}{3}\sin^2\theta_w - \frac{1}{4}\gamma^5$, $\Gamma_{zc}=\frac{1}{4} - \frac{2}{3}\sin^2\theta_w -\frac{1}{4}\gamma^5$ and $q$ is the relative momentum between the two constitute quarks of $(c\bar{b})$-quarkonium. $q_{31}$ and $q_{32}$ are the momenta of the two constitute quarks, i.e.
\begin{equation}
q_{31} = \frac{m_c}{m_{B_c}}{q_3} + q \;\;{\rm and}\;\;
q_{32} = \frac{m_b}{m_{B_c}}{q_3} - q,
\end{equation}
where $m_{B_c}= m_b + m_c$ is implicitly adopted to ensure the gauge invariance of the hard scattering amplitude. $\varepsilon(k)$ is the polarization vector of $Z^0$. $\varepsilon_{s}(p_1)$ and $\varepsilon_{l}(p_1)$ are the polarization vectors relating to the spin and the orbit angular momentum of $(c\bar{b})$- quarkonium, $\varepsilon^{J}_{\mu\nu}(q_3)$ is the polarization tensor for the spin triplet $P$-wave states with $J=0$, $1$ and $2$ respectively. The covariant form of the projectors can be conveniently written as
\begin{equation}
\Pi^0_{q_3}(q)=\frac{-\sqrt{m_{B_c}}}{4{m_b}{m_c}}(\slashed{q}_{32}- m_b) \gamma_5 (\slashed{q}_{31} + m_c),
\end{equation}
and
\begin{equation}
\Pi^\nu_{p_1}(q)=\frac{-\sqrt{m_{B_c}}}{4{m_b}{m_c}}(\slashed{q}_{32}- m_b) \gamma_\nu (\slashed{q}_{31} + m_c),
\end{equation}
After substituting these projectors into the above amplitudes and doing the possible simplifications, the amplitudes then can be squared, summed over the freedoms in the final state and averaged over the ones in the initial state. The selection of the appropriate total angular momentum quantum number is done by performing the proper polarization sum. If defining
\begin{equation}
\Pi_{\alpha\beta}=-g_{\alpha\beta}+\frac{p_{1\alpha} p_{1\beta}}{M^2}\,,
\end{equation}
the sum over polarization for a spin triplet S-state or a spin singlet P-state is given by
\begin{equation}
\sum_{J_z}\varepsilon_\alpha \varepsilon^*_{\alpha'} =\Pi_{\alpha\alpha'} ,\label{3s1}
\end{equation}
where $J_z=s_z$ or $l_z$ respectively. In the case of $^3P_J$ states, the sum over the polarization is given by \cite{projector}
\begin{eqnarray}\label{3pja}
\varepsilon^{(0)}_{\alpha\beta} \varepsilon^{(0)*}_{\alpha'\beta'} &=& \frac{1}{3} \Pi_{\alpha\beta}\Pi_{\alpha'\beta'} \\
\sum_{J_z}\varepsilon^{(1)}_{\alpha\beta} \varepsilon^{(1)*}_{\alpha'\beta'} &=& \frac{1}{2}
(\Pi_{\alpha\alpha'}\Pi_{\beta\beta'}- \Pi_{\alpha\beta'}\Pi_{\alpha'\beta}) \label{3pjb}\\
\sum_{J_z}\varepsilon^{(2)}_{\alpha\beta} \varepsilon^{(2)*}_{\alpha'\beta'} &=& \frac{1}{2}
(\Pi_{\alpha\alpha'}\Pi_{\beta\beta'}+ \Pi_{\alpha\beta'}\Pi_{\alpha'\beta})-\frac{1}{3}
\Pi_{\alpha\beta}\Pi_{\alpha'\beta'} . \label{3pjc}
\end{eqnarray}

We adopt the `new trace amplitude approach' \cite{chang1,tbc2} to derive the analytical expressions for the process $Z^{0}\rightarrow (c\bar{b}) + b +\bar{c}$, where $(c\bar{b})$-quarkonium is in $|(c\bar b)_{\bf 1}[^{1}S_{0}] \rangle$, $|(c\bar b)_{\bf 1}[^{3}S_{1}]\rangle$, $|(c\bar b)_{\bf 1}[^{1}P_{1}]\rangle$, $|(c\bar b)_{\bf 8}[^{1}S_{0}] g\rangle$, $|(c\bar b)_{\bf 1}[^{3}P_{J}]\rangle$ and $|(c\bar b)_{\bf 8}[^{3}S_{1}] g\rangle$ respectively. We have dealt with the case of the two $S$-wave states $|(c\bar b)_{\bf 1}[^{1}S_{0}] \rangle$ and $|(c\bar b)_{\bf 1}[^{3}S_{1}]\rangle$ in Ref.\cite{z0s}. At the present, we continue our work for the $P$-wave states, i.e. to calculate the coefficients together with the independent Lorentz-invariant structures for the $P$-wave case. The derived coefficients are very lengthy and complicated, and to short the paper, we shall not present them here, but shall give the main idea for deriving them in the Appendix \footnote{Detailed formulae can be found in Ref.\cite{coe} and the Mathematica programs in deriving them are available upon request.}.

In doing numerical calculation, we take $m_Z=91.1876$ GeV and $\alpha_s(m_Z)=0.1176$ \cite{pdg}. To be consistent with the present leading-order calculation, we adopt the leading-order $\alpha_s$ running, and by taking the normalization scale to be $2m_c$, which leads to $\alpha_s(2m_c)=0.212$. The two constitute quark masses are taken as $m_b=4.90$ GeV and $m_c=1.50$ GeV. Here, as explained by Ref.\cite{p4}, we take the same (constitute) quark masses for both the $P$-wave and $S$-wave states production. As for the wave function at the origin and the first derivative of the wave function at the origin, we fix their values to be: $|R_S(0)|^2=1.642\; {\rm GeV}^3$ and $|R'_P(0)|^2=0.201\; {\rm GeV}^5$ \cite{pot6}.

By taking the above parameter values, we obtain $\Gamma_{|(c\bar{b})_{\bf 1}[^1S_0]\rangle}=81.4$ KeV, $\Gamma_{|(c\bar{b})_{\bf 1}[^3S_1]\rangle}=116.4$ KeV, $\Gamma_{|(c\bar{b})_{\bf 1}[^1P_1]\rangle}=8.6$ KeV,
$\Gamma_{|(c\bar{b})_{\bf 1}[^3P_0]\rangle}=5.2$ KeV, $\Gamma_{|(c\bar{b})_{\bf 1}[^3P_1]\rangle}=10.5$ KeV,
$\Gamma_{|(c\bar{b})_{\bf 1}[^3P_2]\rangle}=11.6$ KeV, $\Gamma_{|(c\bar{b})_{\bf 8}[^1S_0]\rangle}=10.2 \times{v^4}$ KeV and $\Gamma_{|(c\bar{b})_{\bf 8}[^3S_1]\rangle}=14.5 \times{v^4}$ KeV. It can be found that the decay width of all the $P$-wave states is about $45\%$ of that of the ground state $B_c$ ($|(c\bar{b})_{\bf 1}[^1S_0]\rangle$). Then the $P$-wave states should be taken into consideration so as to make a sound estimation of the $B_c$ meson production. And if taking $v^2 \sim 0.3$, the two color-octet $S$-wave states shall only provide $\sim 3\%$ contribution to the ground state $B_c$.

\begin{figure}
\includegraphics[width=0.40\textwidth]{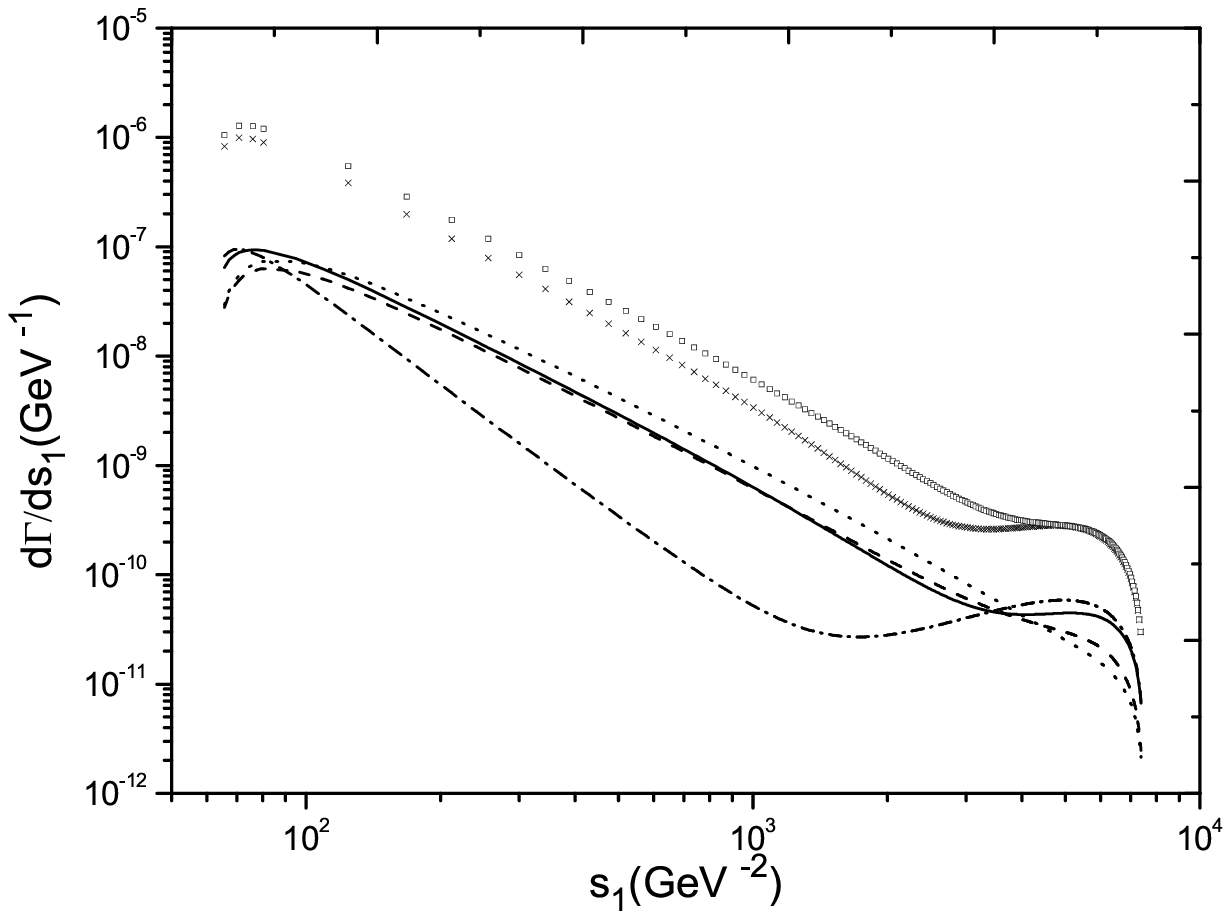}
\includegraphics[width=0.40\textwidth]{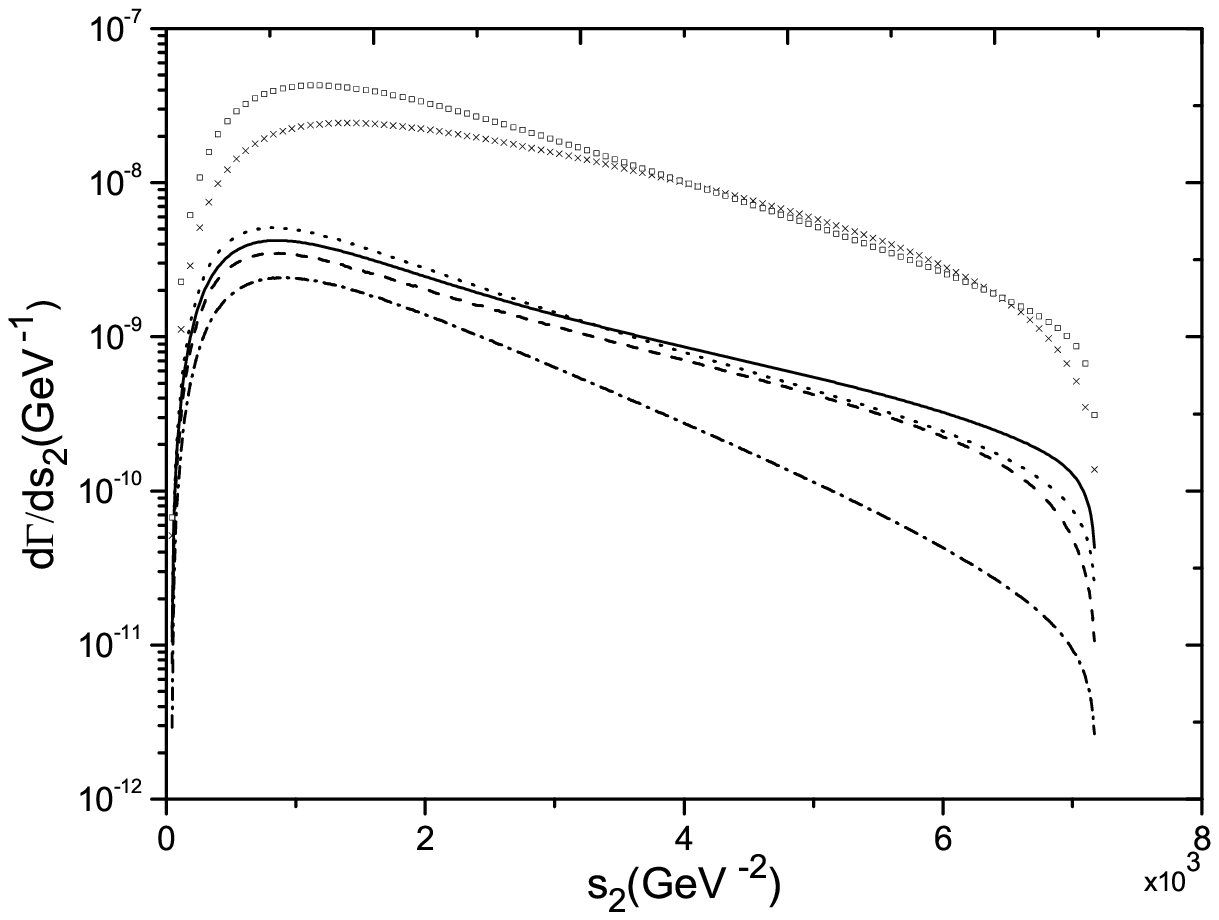}
\caption{Differential decay width $d\Gamma/ds_1$ (Left) and $d\Gamma/ds_2$ (Right) for the process $Z^0\rightarrow (c\bar{b})+b+\bar{c}$, where the squared line, the crossed line, the dotted lines, the solid line, the dashed line and the dash-dot line are for $|(c\bar{b})_{\bf 1}[^3S_1]\rangle$, $|(c\bar{b})_{\bf 1}[^1S_0]\rangle$, $|(c\bar{b})_{\bf 1}[^3P_2]\rangle$, $|(c\bar{b})_{\bf 1}[^3P_1]\rangle$, $|(c\bar{b})_{\bf 1}[^1P_1]\rangle$ and $|(c\bar{b})_{\bf 1}[^3P_0]\rangle$ respectively. } \label{diss1s2}
\end{figure}

\begin{figure}
\includegraphics[width=0.40\textwidth]{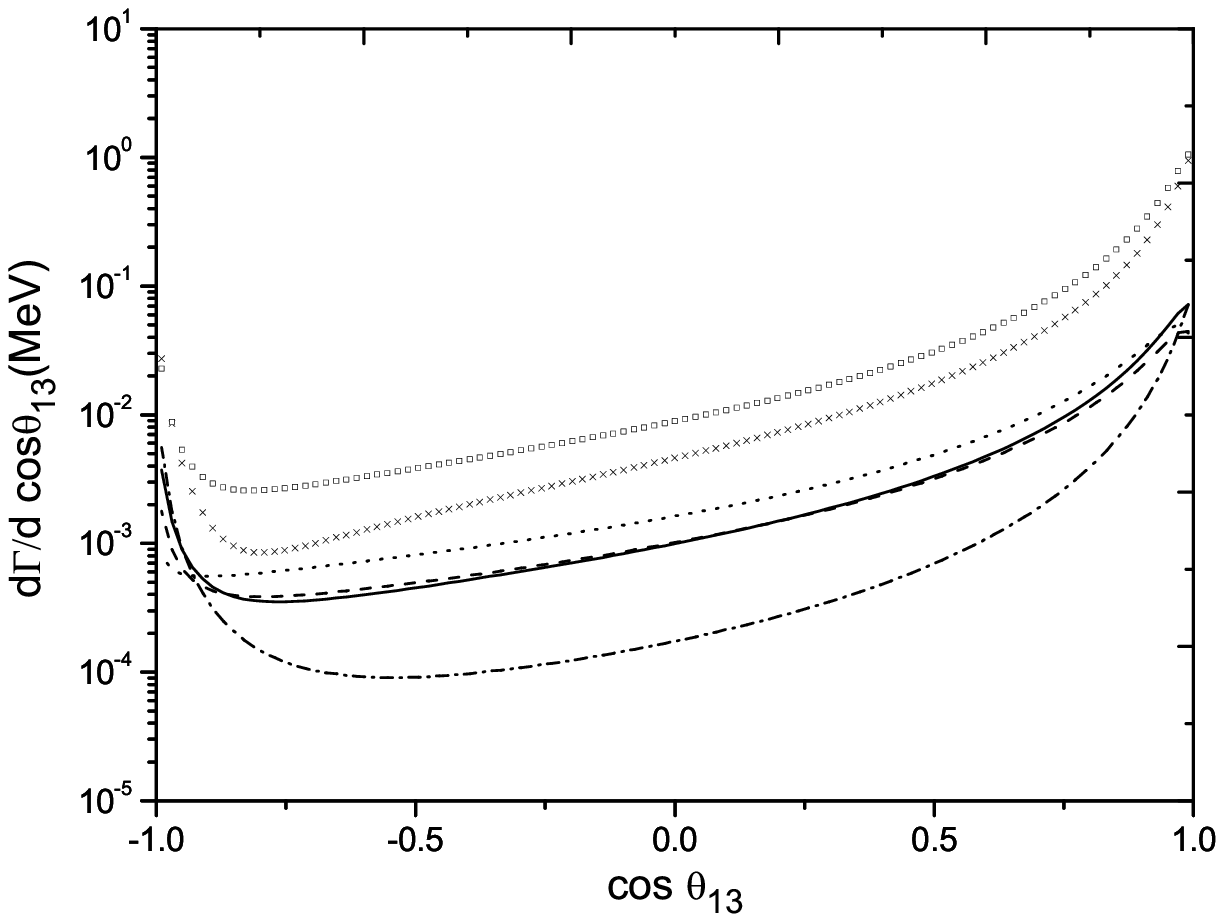}
\hspace{0.2cm}
\includegraphics[width=0.40\textwidth]{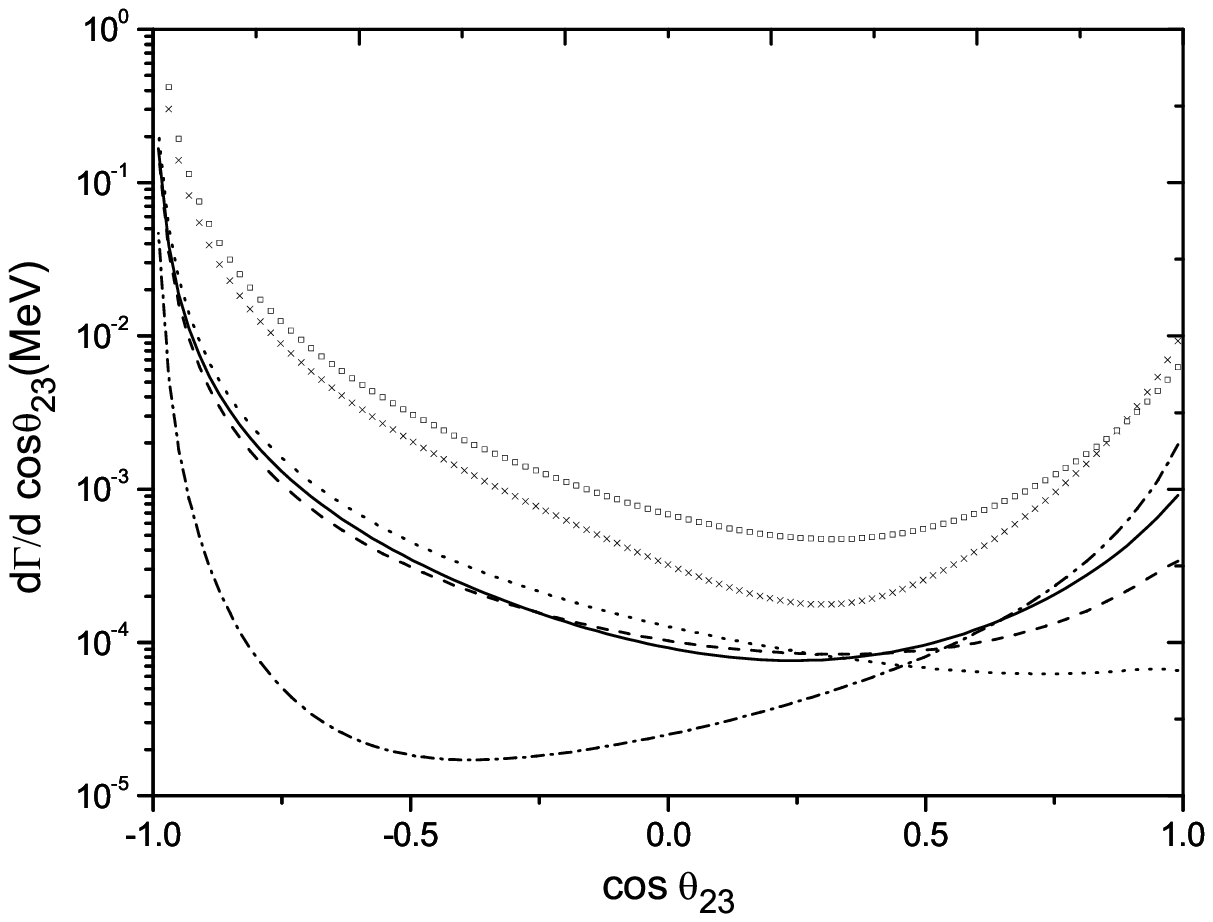}
\caption{Differential decay width $d\Gamma/d\cos{\theta_{13}}$ (Left) and $d\Gamma/d\cos{\theta_{23}}$ (Right) for the process $Z^0\rightarrow (c\bar{b}) +b+\bar c$, where the squared line, the crossed line, the dotted lines, the solid line, the dashed line and the dash-dot line are for $|(c\bar{b})_{\bf 1}[^3S_1]\rangle$, $|(c\bar{b})_{\bf 1}[^1S_0]\rangle$, $|(c\bar{b})_{\bf 1}[^3P_2]\rangle$, $|(c\bar{b})_{\bf 1}[^3P_1]\rangle$, $|(c\bar{b})_{\bf 1}[^1P_1]\rangle$ and $|(c\bar{b})_{\bf 1}[^3P_0]\rangle$ respectively. } \label{discos}
\end{figure}

We present some typical differential distributions for the $(c\bar{b})$-quarkonium production through the $Z^0$ decays in Figs.(\ref{diss1s2},\ref{discos}), where the results for $|(c\bar{b})_{\bf 1}[^3S_1]\rangle$, $|(c\bar{b})_{\bf 1}[^1S_0]\rangle$, $|(c\bar{b})_{\bf 1}[^3P_2]\rangle$, $|(c\bar{b})_{\bf 1}[^3P_1]\rangle$, $|(c\bar{b})_{\bf 1}[^1P_1]\rangle$ and $|(c\bar{b})_{\bf 1}[^3P_0]\rangle$ are presented. Here the two color-singlet $S$-wave states $|(c\bar{b})_{\bf 1}[^3S_1]\rangle$ and $|(c\bar{b})_{\bf 1}[^1S_0]\rangle$ are put together for a comparison. Since the difference between the color-singlet $S$-wave states and the color-octet $S$-wave states is an overall color factor, the shape of their curves are the same, so we do not present the curves for the color-octet ones in Figs.(\ref{diss1s2},\ref{discos}). This is quite different from the case of $B_c$ hadronic production, which has a much more complicated color structure (color flow), and due to the cancelation and enhancement of the different color flows, the color-octet and color-singlet curves behave quite differently \cite{p5}. More explicitly, for the dominant gluon-gluon fusion process ($g+g\to(c\bar{b})+b+\bar{c}$) for the hadornic production, there is five independent color factors for the color-singlet case, while for the color-octet case, the independent color factors change to ten \cite{p5,p6}. This is the reason why the color-octet $S$-wave states can lead to sizable contributions to the hadronic production of $B_c$ in comparison to the color-singlet $P$-wave states \cite{p5}, as is required by the naive NRQCD scaling rules. While for the present case, the color-octet states give somewhat negligible contributions.

Fig.(\ref{diss1s2}) shows the differential distributions of the invariant masses $s_1$ and $s_2$, i.e. $d\Gamma/ds_1$ and $d\Gamma/ds_2$, where $s_1=(q_1+q_3)^2$ and $s_2=(q_1+q_2)^2$. Fig.(\ref{discos}) shows the differential distributions of $\cos\theta_{13}$ and $\cos\theta_{23}$, i.e. $d\Gamma/d\cos\theta_{13}$ and $d\Gamma/d\cos\theta_{23}$, where $\theta_{13}$ is the angle between $\vec{q}_1$ and $\vec{q}_3$, and $\theta_{23}$ is the angle between $\vec{q}_2$ and $\vec{q}_3$. It can be found that for all the considered $(c\bar{b})$-quarkonium states, the largest differential decay width of $d\Gamma/d\cos\theta_{13}$ is achieved when $\theta_{13}=0^{\circ}$, i.e. the $(c\bar{b})$-quarkonium and $c$-quark moving in the same direction. While the largest differential decay width of $d\Gamma/d\cos\theta_{23}$ is achieved when $\theta_{23}=180^{\circ}$, i.e. the $(c\bar{b})$-quarkonium and $b$-quark moving back to back.

Next, it would be interesting to show the theoretical uncertainties for the production. For leading order calculation, its main uncertainty sources include the matrix elements, the renormalization scale $\mu_R$, the quark masses $m_b$ and $m_c$. Since the model-dependent $R_S(0)$ and $R'_P(0)$ emerge as overall factors and their uncertainties can be conveniently discussed when we know their values well, so we shall not discuss such uncertainties in the present paper. $\alpha(\mu_R)$ is another overall parameter, one can set $\mu_R$ to be $2m_c$ or $2m_b$, since the intermediate gluon as shown in Fig.(\ref{feyn}) should be hard enough to produce a $c\bar{c}$-quark pair or a $b\bar{b}$-quark pair. By setting these two scales to calculate the process, we obtain $\Gamma_{\mu_R=2m_b} / \Gamma_{\mu_R=2m_c} \propto \alpha^2_s(2m_b) / \alpha^2_s(2m_c)\sim 0.67$. For definiteness, we fix $\mu_R=2m_c$. As for the uncertainties caused by $m_c$ and $m_b$, we shall study them in `a factorizable way'. When focussing on the uncertainties from $m_c$, we let it be varying within the range of $m_c=1.50\pm0.30$ GeV with all the other factors, including the $b$-quark mass and {\it etc.} being fixed to their center values. Similarly, when discussing the uncertainty caused by $m_b$, we vary the $b$-quark mass $m_b$ within the region of $m_b=4.90\pm0.40$ GeV.

\begin{table}
\begin{tabular}{|c||c|c|c|}
\hline
~~$m_c$({\rm GeV})~~        & ~~1.20~~   & ~~1.50~~   & ~~1.80~~  \\
\hline \hline
$\Gamma_{|(c\bar{b})_{\bf 1}[^1S_0]\rangle}({\rm KeV})$ & 183.5  & 81.4   & 42.2  \\
\hline
$\Gamma_{|(c\bar{b})_{\bf 1}[^3S_1]\rangle}({\rm KeV})$ & 280.1  & 116.4  & 57.1  \\
\hline
$\Gamma_{|(c\bar{b})_{\bf 1}[^1P_1]\rangle}({\rm KeV})$ & 27.6   & 8.6    & 3.2   \\
\hline
$\Gamma_{|(c\bar{b})_{\bf 1}[^3P_0]\rangle}({\rm KeV})$ & 14.7   & 5.2    & 2.3   \\
\hline
$\Gamma_{|(c\bar{b})_{\bf 1}[^3P_1]\rangle}({\rm KeV})$ & 33.8   & 10.5   & 4.1   \\
\hline
$\Gamma_{|(c\bar{b})_{\bf 1}[^3P_2]\rangle}({\rm KeV})$ & 43.8   & 11.6   & 3.9   \\
\hline
\end{tabular}
\caption{Uncertainties for the decay width of the process $Z^0\rightarrow (c\bar{b})+b+\bar{c}$ with varying $m_c$, where $m_b$ is fixed to be $4.9$ GeV. }
\label{tabmc}
\end{table}

\begin{table}
\begin{tabular}{|c||c|c|c|}
\hline ~~$m_b$ ({\rm GeV})~~    & ~~4.50~~   & ~~4.90~~   & ~~5.30~~  \\
\hline \hline
$\Gamma_{|(c\bar{b})_{\bf 1}[^1S_0]\rangle}({\rm KeV})$ & 82.1   & 81.4   & 80.8  \\
\hline
$\Gamma_{|(c\bar{b})_{\bf 1}[^3S_1]\rangle}({\rm KeV})$ & 114.2  & 116.4  & 118.5  \\
\hline
$\Gamma_{|(c\bar{b})_{\bf 1}[^1P_1]\rangle}({\rm KeV})$ & 8.7 & 8.6    & 8.4   \\
\hline
$\Gamma_{|(c\bar{b})_{\bf 1}[^3P_0]\rangle}({\rm KeV})$ & 5.7    & 5.2    & 4.7   \\
\hline
$\Gamma_{|(c\bar{b})_{\bf 1}[^3P_1]\rangle}({\rm KeV})$ & 11.0   & 10.5   & 10.1   \\
\hline
$\Gamma_{|(c\bar{b})_{\bf 1}[^3P_2]\rangle}({\rm KeV})$ & 11.3   & 11.6   & 11.7   \\
\hline
\end{tabular}
\caption{Uncertainties for the decay width of the process $Z^0\rightarrow (c\bar{b})+b+\bar{c}$ with varying $m_b$, where $m_c$ is fixed to be $1.5$ GeV. }
\label{tabmb}
\end{table}

The decay width for the $(c\bar{b})$-quarkonium production through the $Z^0$ decays with varying $m_c$ and $m_b$ are presented in TAB.\ref{tabmc} and TAB.\ref{tabmb}. For the color-singlet $S$-wave states, our present results different from those of Ref.\cite{z0s} at some $m_b$ values as shown by TAB.\ref{tabmb}. It is found that such differences are merely caused by the numerical instability at some singular phase-space points, and at the present, we have improved our numerical treatment so as to make the results more reliable. It shows that the decay width is much more sensitive to $m_c$ than that of $m_b$, and the decay width of the $P$-wave states are more sensitive to quark masses than the $S$-wave states.

By adding the uncertainties caused by $m_b$ and $m_c$ in quadrature, we obtain
\begin{eqnarray}
\Gamma_{|(c\bar{b})_{\bf 1}[^1S_0]\rangle}&=&81.4^{+102.1}_{-39.2} \;{\rm KeV},\\
\Gamma_{|(c\bar{b})_{\bf 1}[^3S_1]\rangle}&=&116.4^{+163.7}_{-59.3} \;{\rm KeV},\\
\Gamma_{|(c\bar{b})_{\bf 1}[^1P_1]\rangle}&=&8.6^{+19.0}_{-5.4} \;{\rm KeV},\\
\Gamma_{|(c\bar{b})_{\bf 1}[^3P_0]\rangle}&=&5.2^{+9.5}_{-2.9} \;{\rm KeV},\\
\Gamma_{|(c\bar{b})_{\bf 1}[^3P_1]\rangle}&=&10.5^{+23.3}_{-6.4} \;{\rm KeV},\\
\Gamma_{|(c\bar{b})_{\bf 1}[^3P_2]\rangle}&=&11.6^{+32.2}_{-7.7} \;{\rm KeV},\\
\Gamma_{|(c\bar{b})_{\bf 8}[^1S_0]g\rangle}&=&10.2^{+12.8}_{-4.9} \times{v^4}  \;{\rm KeV},\\
\Gamma_{|(c\bar{b})_{\bf 8}[^3S_1]g\rangle}&=&14.5^{+20.5}_{-7.4} \times{v^4}  \;{\rm KeV}.
\end{eqnarray}

\begin{figure}
\includegraphics[width=0.4\textwidth]{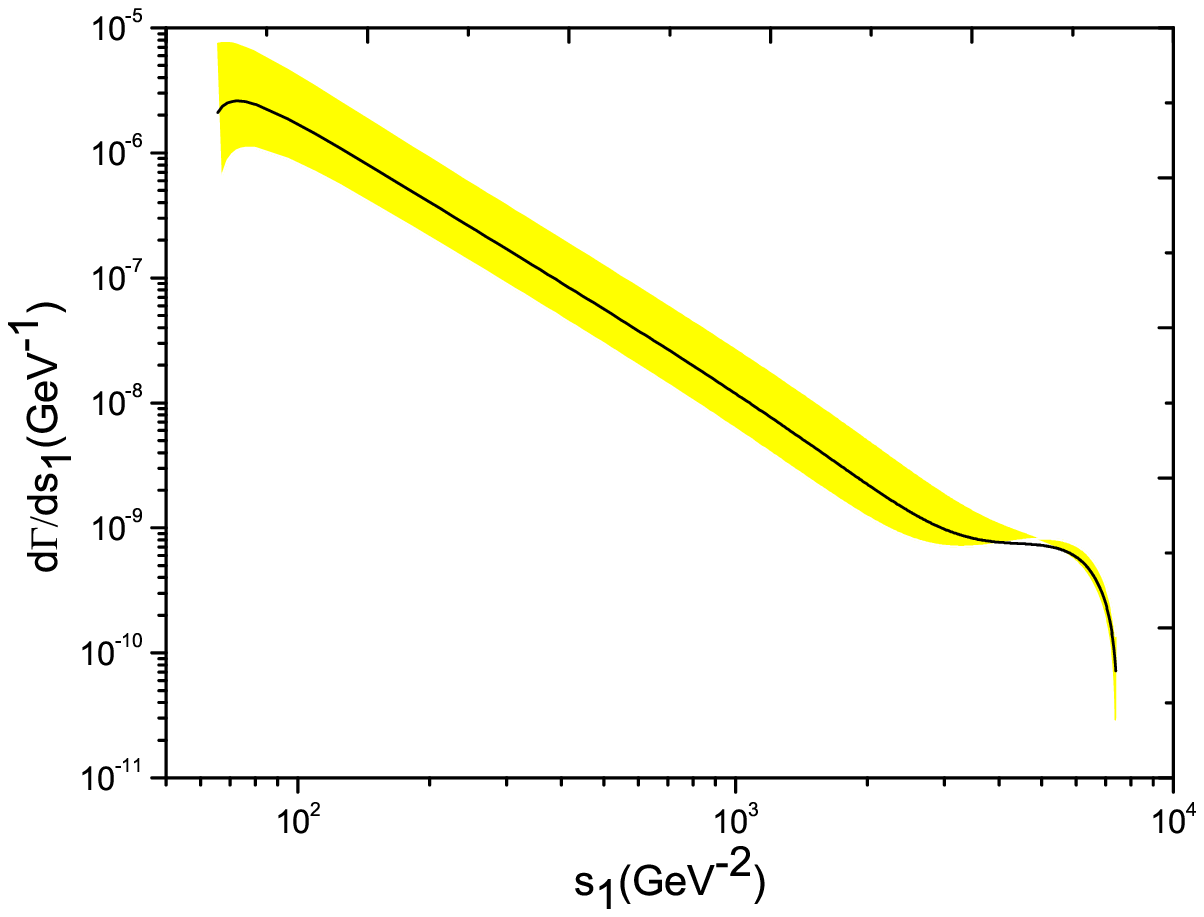}
\hspace{0.2cm}
\includegraphics[width=0.4\textwidth]{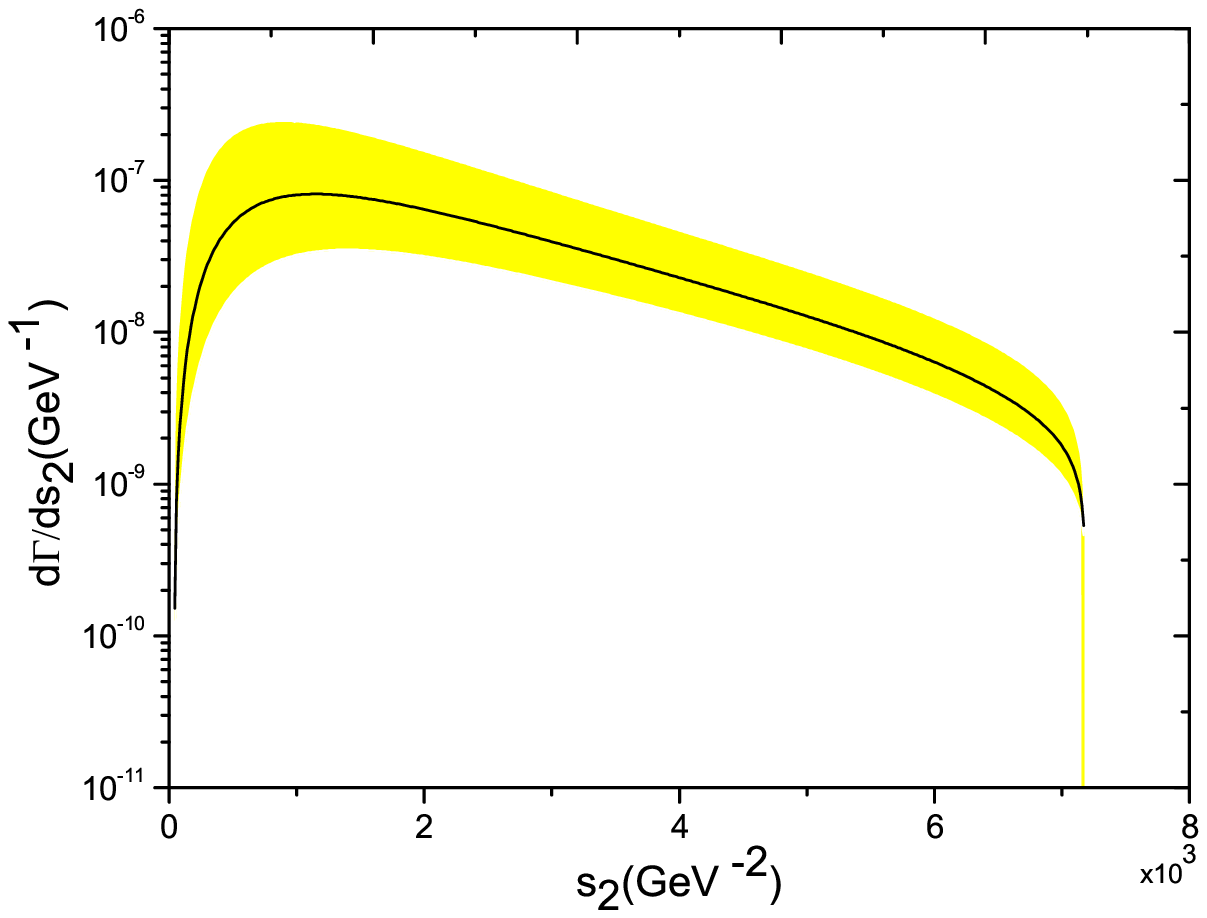}
\caption{Uncertainties of differential decay width $d\Gamma/ds_1$ (Left) and $d\Gamma/ds_2$ (Right) for $Z^0\rightarrow (c\bar{b}) +b+\bar c$, where $|(c\bar{b})_{\bf 1,8}[^3S_1]\rangle$, $|(c\bar{b})_{\bf 1,8}[^1S_0]\rangle$, $|(c\bar{b})_{\bf 1}[^3P_J]\rangle$ and $|(c\bar{b})_{\bf 1}[^1P_1]\rangle$ are taken into consideration. } \label{sun}
\end{figure}

\begin{figure}
\includegraphics[width=0.4\textwidth]{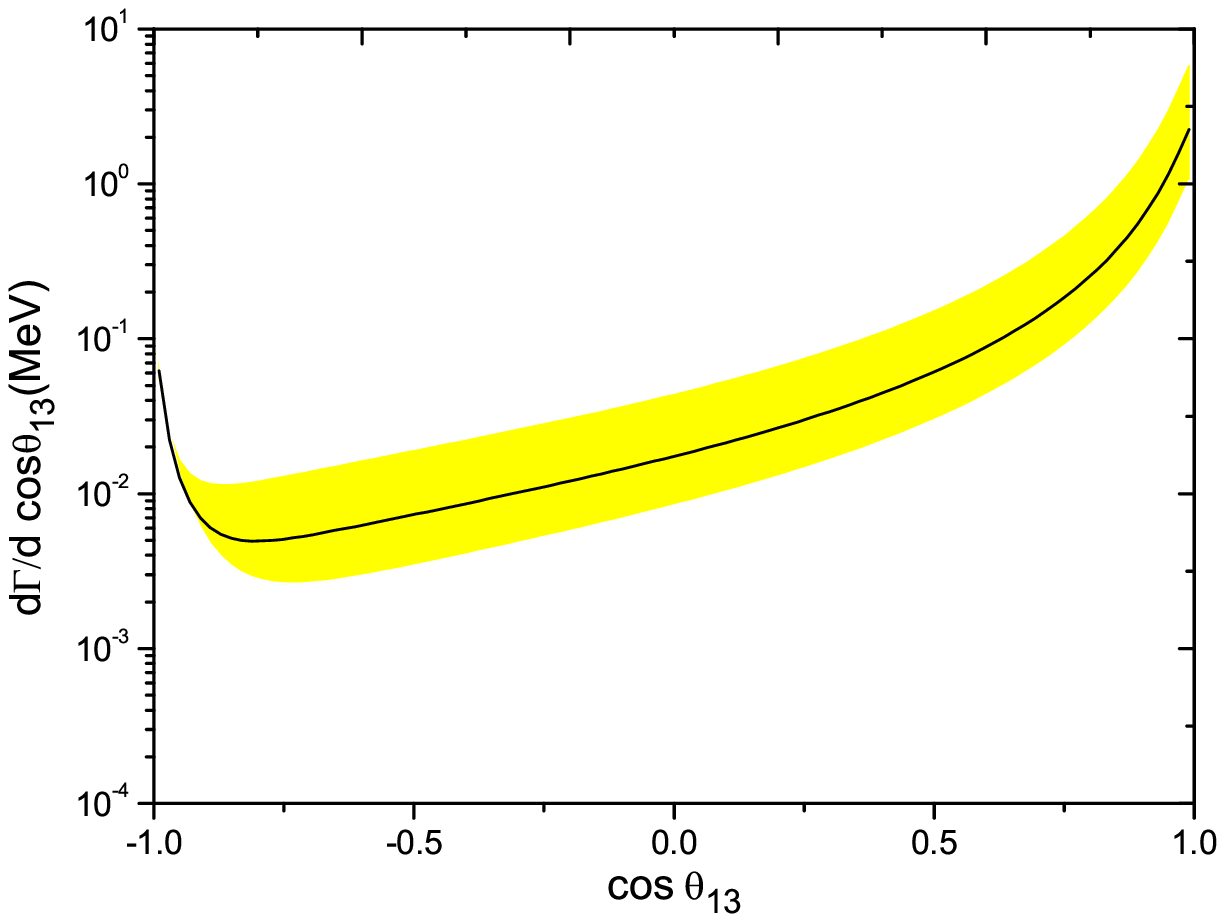}
\hspace{0.2cm}
\includegraphics[width=0.4\textwidth]{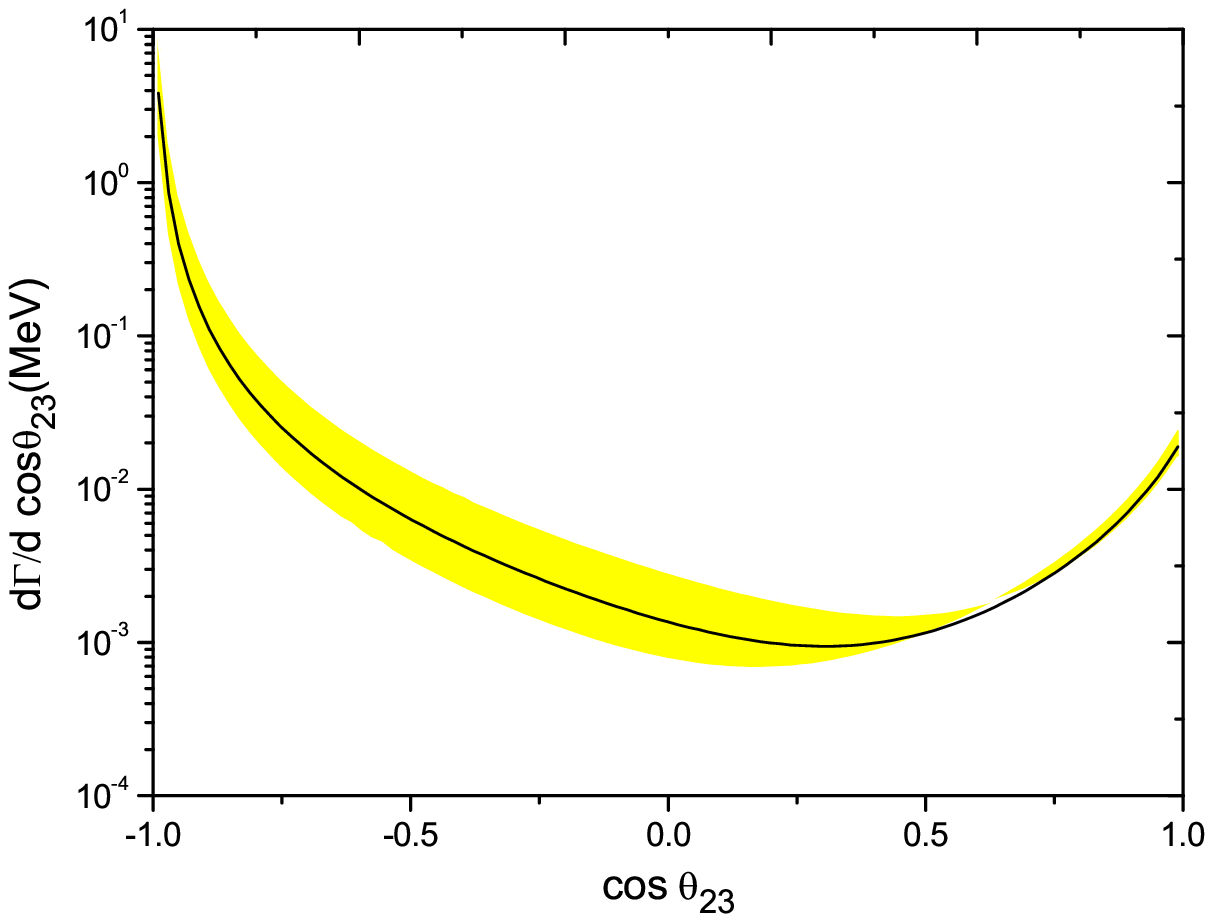}
\caption{Uncertainties of differential decay width $d\Gamma/d\cos{\theta_{13}}$ (Left) and $d\Gamma/d\cos{\theta_{23}}$ (Right) for $Z^0\rightarrow (c\bar{b}) +b+\bar c$, where $|(c\bar{b})_{\bf 1,8}[^3S_1]\rangle$, $|(c\bar{b})_{\bf 1,8}[^1S_0]\rangle$, $|(c\bar{b})_{\bf 1}[^3P_J]\rangle$ and $|(c\bar{b})_{\bf 1}[^1P_1]\rangle$ are taken into consideration. } \label{cosun}
\end{figure}

If assuming the higher excited states decay to the ground state $B_c$($|(c\bar{b})_{\bf 1}[^1S_0]\rangle$) with $100\%$ efficiency, then we obtain the total decay width for the process $Z^0 \rightarrow B_{c} +b +\bar{c}$, i.e.
\begin{eqnarray}
\Gamma_{Z^0\to B_c}&=&235.9^{+352.8}_{-122.0} \;{\rm KeV}\;,
\end{eqnarray}
where for the color-octet $S$-wave matrix element, we take $v^2=0.20$ as its central value. It is found that the total decay width of all the $P$-wave states, including the two color-octet $S$-wave quarkonium states' contributions, is around $20\%$ of the total decay width. So these higher Fock states contributions should be taken into consideration, especially for the future high collision and high luminosity colliders. The shaded bands in Figs.(\ref{sun},\ref{cosun}) show the corresponding uncertainty more clearly, where all the above mentioned Fock states $|(c\bar{b})_{\bf 1,8}[^1S_0]\rangle$, $|(c\bar{b})_{\bf 1,8}[^3S_1]\rangle$, $|(c\bar{b})_{\bf 1}[^1P_1]\rangle$ and $|(c\bar{b})_{\bf 1}[^3P_J]\rangle$ are summed up. The center solid line is for $m_c=1.5 {\rm GeV}$ and $m_b=4.9 {\rm GeV}$, the upper edge of the band is obtained by setting $m_c=1.2 {\rm GeV}$ and $m_b=5.3 {\rm GeV}$, while the lower edge of the band is obtained by setting $m_c=1.8 {\rm GeV}$ and $m_b=4.5 {\rm GeV}$.

As a summary: In the present paper, we have finished a discussion on the $(c\bar{b})$-quarkonium production through the $Z^0$ decays, where $(c\bar{b})$-quarkonium in $|(c\bar{b})_{\bf 1,8}[^3S_1]\rangle$, $|(c\bar{b})_{\bf 1,8}[^1S_0]\rangle$, $|(c\bar{b})_{\bf 1}[^3P_J]\rangle$ and $|(c\bar{b})_{\bf 1}[^1P_1]\rangle$ have been taken into consideration. We have adopted the `New Trace Technology' to obtain the analytic expressions for the amplitude, which shall be helpful for certain cases, especially when one wants to simulate the $B_c$ events at the suggested $Z$-factory \cite{gigaz,wjw}.

It is found that the $P$-wave states can provide sizable contributions to the ground state ($B_c$) production, which is about $45\%$ of that of the ground state and leads to $\sim 20\%$ contribution to the total decay width. Under such a comparatively large production rate from $Z^{0}$ decays, the $P$-wave quarkonium itself is worthwhile to study the possibility of directly measuring the $P$-wave states, which is very important in understanding the $(c\bar{b})$-quarkonium mass spectrum and testing the potential models. It is found that the $P$-wave states are more sensitive to the quark masses than the $S$-wave states. If all the low-laying excited states decay to the ground state $B_c ((c\bar{b})_{\bf 1}[^1S_0])$, we obtain the total decay width for $B_c$ production, i.e. $235.9^{+352.8}_{-122.0}$ KeV, where the errors are caused by varying $m_b$ and $m_c$ within their reasonable regions $m_c\in[1.2, 1.8]$ GeV and $m_c\in[4.5,5.3]$ GeV and by varying $v^2\in(0.1,0.3)$.

\hspace{2cm}

{\bf Acknowledgements}: This work was supported in part by the Fundamental Research Funds for the Central Universities under Grant No.CDJXS11100005, by Natural Science Foundation Project of CQ CSTC under Grant No.2008BB0298, by Natural Science Foundation of China under Grant No.10805082 and No.11075225. \\

\appendix

\section*{Appendix: Main idea for deriving the amplitude of $Z^0(k)\rightarrow (c\bar{b})(q_3) + b(q_2) +\bar{c}(q_1)$}

We adopt the `new trace amplitude approach' \cite{chang1,tbc2} to derive the analytical expression for $Z^{0}\rightarrow (c\bar{b}) + b +\bar{c}$, where $(c\bar{b})$-quarkonium is in $|(c\bar b)_{\bf 1}[^{1}S_{0}] \rangle$, $|(c\bar b)_{\bf 1}[^{3}S_{1}]\rangle$, $|(c\bar b)_{\bf 1}[^{1}P_{1}]\rangle$, $|(c\bar b)_{\bf 8}[^{1}S_{0}] g\rangle$, $|(c\bar b)_{\bf 1}[^{3}P_{J}]\rangle$ and $|(c\bar b)_{\bf 8}[^{3}S_{1}] g\rangle$. The detailed formulae for the present case can be found in Appendix B of Ref.\cite{z0s}. Here, we shall only present the main idea of the approach.

As shown by Fig.(\ref{feyn}), there are four Feynman diagrams (amplitudes), we first arrange each of the four amplitudes listed in Eqs.(\ref{A1},\ref{A2},\ref{A3},\ref{A4}) or Eqs.(\ref{A5},\ref{A6},\ref{A7},\ref{A8}) into four orthogonal sub-amplitudes according to the four spin combinations of the outgoing $b$-quark and $\bar{c}$-antiquark. Next, we do the trace of the Dirac-$\gamma$ matrix strings at the amplitude level by properly dealing with the massive spinors, which will result in explicit series over independent Lorentz-invariant structures. Finally, we determine the analytical expressions for the coefficients of these Lorentz-invariant structures.

The independent coefficients can be schematically represented by $A^i_j$, where $i=(1,2,3',4')$ and $j=(1,\cdots,n)$ with $n$ equals to the maximum independent Lorentz structure number for a particular quarkonium state. The coefficients of the color-singlet $S$-wave states $|(c\bar b)_{\bf 1}[^{1}S_{0}] \rangle$ and $|(c\bar b)_{\bf 1}[^{3}S_{1}]\rangle$ have been presented in Ref.\cite{z0s}. The results for the color-octet $S$-wave states, $|(c\bar b)_{\bf 8}[^{1}S_{0}] g\rangle$ and $|(c\bar b)_{\bf 8}[^{3}S_{1}] g\rangle$, can be easily read from the color-singlet $S$-wave case, since the only difference is an overall color factor. For spin-singlet $|(c\bar{b})_{\bf 1}[^1P_1]\rangle$, there are twelve basic lorentz structures, which are the same as that of $|(c\bar{b})_{\bf 1}[^3S_1]\rangle$ \cite{z0s}, only one need to change the spin polarization vector there to the present orbit polarization vector. For the spin-triplet $|(c\bar{b})_{\bf 1}[^3P_J]\rangle$ states, there are totally 34 independent basic Lorentz structures $B_j$, i.e.
\begin{widetext}
\begin{eqnarray}
B_1&=& \frac{1}{m_Z}q_{2}\cdot\epsilon(k)\varepsilon^J_{\alpha\alpha},\;
B_2 =\frac{1}{m_Z}q_{3}\cdot\epsilon(k)\varepsilon^J_{\alpha\alpha},\;
B_3 =\frac{1}{m_Z}q_{2\alpha}\epsilon_\beta(k)\varepsilon^J_{\alpha\beta},\;
B_4 = \frac{1}{m_Z}k_{\alpha}\epsilon_\beta(k)\varepsilon^J_{\alpha\beta},\nonumber\\
B_5&=&\frac{i\varepsilon^J_{\alpha\alpha}}{m_Z^3}\varepsilon(k,q_2,q_3,\epsilon(k)),\;
B_6 =\frac{i\varepsilon^J_{\alpha\beta}}{m_Z^3}\varepsilon(k,q_2,q_3,\alpha)\epsilon_\beta(k),\;
B_7 = \frac{i\varepsilon^J_{\alpha\beta}}{m_Z^3}\varepsilon(k,q_3,\epsilon(k),\alpha)q_{2\beta} ,\nonumber\\
B_8 &=&\frac{i\varepsilon^J_{\alpha\beta}}{m_Z^3}\varepsilon(k,q_2,\alpha,\beta)q_{2}\cdot\epsilon(k),\;
B_9 = \frac{i\varepsilon^J_{\alpha\beta}}{m_Z^3}\varepsilon(k,q_3,\epsilon(k),\alpha)k_{\beta},\;
B_{10}=\frac{i\varepsilon^J_{\alpha\beta}}{m_Z^3}\varepsilon(k,q_2,\epsilon(k),\alpha)k_{\beta} ,\nonumber\\
B_{11} &=& \frac{i\varepsilon^J_{\alpha\beta}}{m_Z^3}\varepsilon(k,q_3,\alpha,\beta)q_{3}\cdot\epsilon(k),\;
B_{12}=\frac{i\varepsilon^J_{\alpha\beta}}{m_Z^3}\varepsilon(k,q_2,\alpha,\beta)q_{3}\cdot\epsilon(k),\;
B_{13} = \frac{i\varepsilon^J_{\alpha\beta}}{m_Z^3}\varepsilon(k,q_3,\alpha,\beta)q_{2}\cdot\epsilon(k),\nonumber\\
B_{14} &=&\frac{i\varepsilon^J_{\alpha\beta}}{m_Z^3}\varepsilon(k,q_2,\epsilon(k),\alpha)q_{2\beta},\;
B_{15} = \frac{i\varepsilon^J_{\alpha\beta}}{m_Z^3}\varepsilon(q_2,q_3,\epsilon(k),\alpha)q_{2\beta},\;
B_{16}=\frac{i\varepsilon^J_{\alpha\beta}}{m_Z^3}\varepsilon(q_2,q_3,\alpha,\beta)q_{2}\cdot\epsilon(k),\nonumber\\
B_{17} &=& \frac{i\varepsilon^J_{\alpha\beta}}{m_Z^3}\varepsilon(q_2,q_3,\alpha,\beta)q_{3}\cdot\epsilon(k),\;
B_{18} = \frac{i\varepsilon^J_{\alpha\beta}}{m_Z^3}\varepsilon(q_2,q_3,\epsilon(k),\alpha)k_{\beta},\;
B_{19} =\frac{i\varepsilon^J_{\alpha\beta}}{m_Z}\varepsilon(k,\epsilon(k),\alpha,\beta),\nonumber\\
B_{20} &=&\frac{i\varepsilon^J_{\alpha\beta}}{m_Z}\varepsilon(q_2,\epsilon(k),\alpha,\beta),\;
B_{21} =\frac{i\varepsilon^J_{\alpha\beta}}{m_Z}\varepsilon(q_3,\epsilon(k),\alpha,\beta),\;
B_{22}  = \frac{i\varepsilon^J_{\alpha\beta}}{m_Z^5}\varepsilon(k,q_2,q_3,\alpha)k_{\beta} q_3\cdot\epsilon(k),\nonumber\\
B_{23} &=&\frac{i\varepsilon^J_{\alpha\beta}}{m_Z^5}\varepsilon(k,q_2,q_3,\alpha)q_{2\beta} q_2\cdot\epsilon(k),\;
B_{24}=\frac{i\varepsilon^J_{\alpha\beta}}{m_Z^5}\varepsilon(k,q_2,q_3,\alpha)k_{\beta} q_2\cdot\epsilon(k),\;
B_{25} =\frac{i\varepsilon^J_{\alpha\beta}}{m_Z^5}\varepsilon(k,q_2,q_3,\alpha)q_{2\beta} q_3\cdot\epsilon(k),\nonumber\\
B_{26}  &=& \frac{i\varepsilon^J_{\alpha\beta}}{m_Z^5}\varepsilon(k,q_2,q_3,\epsilon(k))q_{2\alpha} q_{2\beta},\;
B_{27} =\frac{i\varepsilon^J_{\alpha\beta}}{m_Z^5}\varepsilon(k,q_2,q_3,\epsilon(k))k_{\alpha} k_{\beta},\;
B_{28} = \frac{i\varepsilon^J_{\alpha\beta}}{m_Z^5}\varepsilon(k,q_2,q_3,\epsilon(k))k_{\alpha} q_{2\beta},\nonumber\\
B_{29} &=&\frac{\varepsilon^J_{\alpha\beta}}{m_Z^3}k_{\alpha}k_{\beta}q_{3}\cdot\epsilon(k),\;
B_{30} = \frac{\varepsilon^J_{\alpha\beta}}{m_Z^3}k_{\alpha}q_{2\beta}q_{3}\cdot\epsilon(k),\;
B_{31} =\frac{\varepsilon^J_{\alpha\beta}}{m_Z^3}k_{\alpha}k_{\beta}q_{2}\cdot\epsilon(k),\nonumber\\
B_{32} &=&\frac{\varepsilon^J_{\alpha\beta}}{m_Z^3}k_{\alpha}q_{2\beta}q_{2}\cdot\epsilon(k),\;
B_{33} = \frac{\varepsilon^J_{\alpha\beta}}{m_Z^3}q_{2\alpha}q_{2\beta}q_{3}\cdot\epsilon(k),\;
B_{34}=\frac{\varepsilon^J_{\alpha\beta}}{m_Z^3}q_{2\alpha}q_{2\beta}q_{2}\cdot\epsilon(k).
\end{eqnarray}
\end{widetext}

It is noted that $\varepsilon^{0,2}_{\alpha\beta}$ is the symmetric tensor and $\varepsilon^{1}_{\alpha\beta}$ is the anti-symmetric tensor, and the fact that $\varepsilon^{1}_{\alpha\alpha}=\varepsilon^{2}_{\alpha\alpha}=0$. so the terms involving the following coefficients have no contributions to the square of the amplitude, and practically, we can safely set the coefficients before them to be zero:
\begin{equation}
\begin{array}{c}
A^i_j(|(c\bar{b})_{\bf 1}[^3P_0]\rangle)=0 \;\;\;\;\;\;\;\;{\rm for}\;i=(1-4),\\
j=(8,9,11,12,13,15,16,17,19,20,21,22,23,24,25)\\
\end{array}
\end{equation}
\begin{equation}
\begin{array}{c}
A^i_j(|(c\bar{b})_{\bf 1}[^3P_1]\rangle)=0 \;\;\;\;\;\;\;\;{\rm for}\;i=(1-4), \\
j=(1,2,5,26,27,29,31,33,34)\\
\end{array}
\end{equation}
\begin{equation}
\begin{array}{c}
A^i_j(|(c\bar{b})_{\bf 1}[^3P_2]\rangle)=0 \;\;\;\;\;\;\;\;{\rm for}\;i=(1-4),\\
j=(1,2,5,8,11,12,13,16,17,19,20,21) .
\end{array}
\end{equation}
The non-zero coefficients for the independent Lorentz-invariant structures are very lengthy and complicated, so to short the paper, we shall not present them here. The interesting reader can turn to Ref.\cite{coe} for detailed expressions.

\end{document}